\pdfoutput=1

\documentclass[lettersize,journal]{IEEEtran}

\usepackage{amsmath,amsfonts}
\usepackage{algorithm}
\usepackage{algpseudocode}
\usepackage{array}
\usepackage[caption=false,font=normalsize,labelfont=sf,textfont=sf]{subfig}
\usepackage{textcomp}
\usepackage{stfloats}
\usepackage{url}
\usepackage{verbatim}
\usepackage{graphicx}    
\usepackage{cite}
\usepackage{braket}
\usepackage{booktabs}

\DeclareGraphicsExtensions{.pdf,.png,.jpg}

\begin{document}

\title{Dynamic-Key–Aware Co-Simulation Framework for Next Generation of SCADA Systems Encrypted by Quantum-Key-Distribution Techniques}

\author{Ziqing Zhu,~\IEEEmembership{Member,~IEEE} 
}
\maketitle

\begin{abstract}
To address growing cybersecurity challenges in modern power dispatch systems, this paper proposes a multi-layer modeling and optimization framework for SCADA systems enhanced with quantum key distribution (QKD). While most existing applications of QKD in the power sector focus on building secure point-to-point communication tunnels, they rarely consider the system-level coupling between key dynamics and control scheduling. In contrast, our approach integrates quantum key generation, consumption, inventory prediction, and control latency into a unified model, enabling key-aware reconfiguration of SCADA control chains based on task security demands and real-time resource constraints. To resolve conflicts in key resource allocation between transmission system operators (TSOs) and distribution system operators (DSOs), we formulate a bi-level Stackelberg game and transform it into a mathematical program with complementarity constraints (MPCC). We further develop an efficient Level Decomposition–Complementarity Pruning (LD-CP) algorithm to solve the problem. To support reproducible evaluation, we build an end-to-end co-simulation platform that integrates physical-layer disruptions via OpenQKD-Sim, Q3P/IEC-104 protocol stack binding, and real-time control-chain monitoring through Grafana. Experimental results on the IEEE 39- and 118-bus systems show that our method increases task success rate by 25\%, reduces peak frequency deviation by 70\%, and improves key utilization to 83\%. This work lays the foundation for future quantum-secure control systems in power grid operations.
\end{abstract}

\section{Introduction}
\subsection{Vision and Significance of QKD-Enabled SCADA}
\IEEEPARstart{I}{n} modern power systems, the evolution of SCADA is driven not only by the need for ultra-low-latency control but also by the rising demand for long-term cybersecurity. Quantum Key Distribution (QKD) has emerged as the only known technology that guarantees information-theoretic security, making it an ideal foundation for next-generation control infrastructures [1]. By enabling unbreakable encryption for critical control signals, such as frequency regulation or emergency load shedding, QKD eliminates the risk of key compromise, even in the face of quantum adversaries [2]. Moreover, its property of forward secrecy simplifies key lifecycle management by removing reliance on certificate hierarchies or algorithmic assumptions [3]. Beyond encryption, the physical randomness embedded in QKD channels can also support anomaly detection mechanisms at the communication layer [4]. Together, these features make QKD not just a technological upgrade, but a strategic enabler of resilient and future-proof SCADA systems.

\subsection{Related Work and Research Gap}

\subsubsection{Prototype QKD Deployments in Power Grids}
Several pilot projects have applied quantum key distribution (QKD) to secure power system communications. For example, ID Quantique and KEPCO deployed one of the first QKD-protected substation communication links in Korea, using QKD over a 40 km optical ground wire between two substations [5]. This 2021 prototype demonstrated basic point-to-point encryption for substation SCADA traffic. Similarly, State Grid Corporation of China (SGCC) has conducted trial installations of QKD for substation and distribution network communications. In one demonstration, a multi-relay QKD network spanned three substations and secured remote telemetry/command exchanges in a provincial dispatch grid [6]. These projects proved that QKD can be integrated into utility fiber infrastructure, but they focused on connectivity and basic functionality. Notably, none provided a comprehensive system-level performance analysis – metrics like end-to-end latency impact, key consumption rates under power control traffic patterns, or reliability of protection signaling with QKD were not rigorously evaluated.

\subsubsection{Resource-Aware Key Management in QKD Networks}
In the telecommunications domain, researchers have explored adaptive key management to mitigate QKD’s limited key generation rates. Strategies such as maintaining quantum key pools (buffers of QKD-generated keys) and hybrid encryption switching have been proposed. In these schemes, high-security one-time-pad (OTP) encryption is used when key supply is abundant, and the system falls back to conventional symmetric ciphers (e.g. AES) when keys are scarce [7]. This hybrid approach conserves one-time key material by using computationally secure AES encryption as needed, at the cost of reducing the theoretical security to conventional levels [8]. While such resource-aware schemes were designed for general QKD networks, they ignore the hard real-time constraints of power-system control loops. Power grid control messages (protective tripping commands, load control signals, etc.) often have strict latency deadlines (on the order of milliseconds to tens of milliseconds). A key-management strategy that withholds data waiting for fresh keys or that switches encryption modes on-the-fly could violate these timing requirements. For instance, using a slower key update or allowing key exhaustion to introduce a 100+ ms delay can cause grid synchronization errors [9]. Prior works on QKD key management did not account for these real-time performance needs unique to energy systems.

\subsubsection{Hierarchical TSO–DSO Optimization Approaches}
There is rich literature on coordinating operations between Transmission System Operators (TSOs) and Distribution System Operators (DSOs) in a smart grid. Many works formulate hierarchical optimization or game-theoretic models to align transmission-level dispatch with distribution-level resources [10-12]. Yet, none of these studies consider the integration of QKD-based security constraints. The decision frameworks assume control signals can be issued securely without limitation. In reality, QKD keys are a stochastic and non-duplicable resource – key generation rates may fluctuate with quantum channel conditions, and each key bit can only be used once. No existing TSO–DSO optimization incorporates a coupling between control actions and available secure key supply. For example, scheduling a fast DER control response might be infeasible if encryption keys are insufficient at that moment, but current models do not capture such coupling. Thus, the impact of limited key availability on hierarchical control decisions remains an open question.

\subsubsection{Simulation Toolchains for Cyber–Physical–Quantum Systems}
To date, studies of power grid QKD integration have used separate simulation tools for different domains. Communication network simulators (such as OPNET, ns-3, or OMNeT++) and power system simulators (such as OPAL-RT real-time digital simulators or PSCAD/EMTDC) are well established. Co-simulation platforms exist to combine power and communication models, enabling analysis of cyber-physical interactions in smart grids [13]. For example, a real-time testbed may use OPAL-RT to simulate the electrical grid and ns-3 or OMNeT++ to simulate the ICT network, synchronized to exchange data in each time step [14]. On the quantum side, specialized tools like OpenQKD-Sim (a module extending ns-3 for QKD networks) have been developed to simulate QKD link physics and key distribution processes [15]. However, an integrated test environment that combines all three layers – power system dynamics, communication networking, and quantum key generation – is still missing. Most studies either evaluate communication/security aspects in isolation (using a network or QKD simulator alone) or examine grid control impacts assuming an abstracted communications layer. There is a lack of end-to-end co-simulation platforms or testbeds that can validate how QKD-secured control schemes perform under realistic grid conditions, including the delays of networking and the probabilistic behavior of quantum key generation.

\subsection{Research Gap and Main Contributions}
In summary, few existing framework provides a holistic solution that (i) couples the dynamic availability of QKD keys with grid control actions across multiple timescales, (ii) enables fair and efficient key allocation between transmission and distribution operators in a coordinated manner, and (iii) validates these concepts on an end-to-end cyber–physical–quantum co-simulation or prototype platform. 

Correspondingly, this work addresses the integration of QKD into real-time power system operations by establishing a comprehensive framework that spans from physical-layer key generation to high-level control optimization. The main contributions are summarized as follows:

(1) We develop the first unified multi-layer model that explicitly links the stochastic generation of QKD keys with their consumption across control messages, dynamically tracks key inventory, and models the resulting impact on control-chain latency. This framework enables a realistic assessment of how QKD-induced constraints propagate through secure SCADA operations, bridging the gap between quantum communication and cyber-physical control.

(2) We propose a novel chance-constrained optimization formulation for key-aware control-chain reconfiguration. By encoding both communication reliability and key availability into a mixed-integer probabilistic scheduling model, we ensure safety-critical control loops are resilient under key scarcity. We further extend this model to a hierarchical transmission–distribution coordination setting using a bilevel Stackelberg game formulation. To efficiently solve this nonconvex Mathematical Program with Equilibrium Constraints (MPEC), we design a Level Decomposition with Complementarity Pruning (LD-CP) algorithm that avoids full enumeration of complementarity pairs and converges to M-stationary solutions with strong practical scalability.

(3) We implement a modular, end-to-end testbed for QKD-enhanced SCADA co-simulation. The platform integrates (i) OpenQKD-Sim for modeling link-level photon loss and channel outages, (ii) a hybrid protocol stack combining Q3P and IEC 60870-5-104 for secure control signaling, and (iii) Grafana-based real-time visualization and configuration tools. The platform supports extensions for hardware-in-the-loop (HIL) experiments via RTDS and is designed for portability and reproducibility.

\subsection{Paper Structure}
The remainder of this paper is organized as follows. Section 2 introduces the multi-layer architecture of the QKD-enhanced SCADA system and formulates the core constraints on key generation, inventory dynamics, and control latency. Section 3 presents the key-aware control-chain reconfiguration problem and its Stackelberg extension for TSO–DSO coordination, along with the proposed LD-CP algorithm. Section 4 describes the implementation of our modular co-simulation testbed, including quantum link modeling, protocol binding, and visualization components. Section 5 details the experimental setup and performance results on IEEE benchmark systems. Finally, Section 6 concludes the paper and outlines future directions.

\section{Dynamic Key Resource Modeling}

\subsection{State Variables and Time Scale}

In building a QKD-SCADA co-simulation platform for power dispatch, a key challenge is modeling the dynamic interplay of key generation, usage, and inventory. Unlike conventional systems limited by bandwidth, QKD generates non-replicable, non-bufferable keys whose availability is tightly linked to physical conditions. To capture this, we define the following continuous-time state variables for quantum key behavior. Let \( t \) denote the continuous time in seconds. The variable \( K(t) \in \mathbb{R}_+ \) represents the available quantum key inventory (in bits) at time \( t \), corresponding to the usable key pool maintained at the dispatch center. The function \( G(t) \) denotes the instantaneous key generation rate (bits per second), as determined by the underlying QKD link. For each class of control task indexed by \( i \in \{1, \dots, T\} \), we define \( C_i(t) \) as the instantaneous key consumption rate (bits per second). The total number of distinct control task types, denoted by \( T \), typically includes wide-area monitoring (e.g., PMU data streams) and active control commands (e.g., AGC setpoints, AVR signals, load shedding instructions), which operate across different timescales in power system operation. For numerical implementation, we adopt a fixed sampling interval \( \Delta t \), and define discrete-time representations such as \( K_k = K(k \Delta t) \), \( G_k = G(k \Delta t) \), and \( C_{i,k} = C_i(k \Delta t) \) for each time step \( k \in \mathbb{N} \).

\subsection{Key Generation Model}

The key generation rate is a critical metric in QKD systems, directly affecting how fast the key pool is replenished. In power systems, QKD links are often deployed over OPGW or dedicated optical fibers, whose performance is sensitive to environmental factors (e.g., wind, icing, temperature) and disturbances from power equipment. For BB84-based QKD systems, the instantaneous key generation rate is given by:
\begin{equation}
G(t) = \underbrace{R_p \cdot \eta(t)}_{\text{photon detection rate}} \cdot 
       \underbrace{q(t)}_{\text{sifted-bit retention ratio}} \cdot 
       f_{\text{sec}}\!\left(\text{QBER}(t)\right)
\end{equation}
where \( R_p \) denotes the photon emission rate at the sender (in photons per second), determined by the QKD device configuration. The function \( \eta(t) \in [0,1] \) represents the instantaneous transmission efficiency of the link, which reflects the proportion of photons successfully transmitted from the source to the detector. This efficiency is affected by factors such as optical fiber attenuation, splicing loss, wind-induced sagging, and temperature-induced refraction changes. The term \( q(t) \) denotes the sifting ratio, i.e., the proportion of measurement outcomes retained after basis reconciliation. The quantity \( \text{QBER}(t) \) stands for the quantum bit error rate at time \( t \), which increases under electromagnetic interference, misalignment, or vibration, commonly present in substation environments. The function \( f_{\text{sec}}(\cdot) \) denotes the \emph{privacy amplification factor}, which quantifies the fraction of sifted bits that can be securely retained as final key material. It is typically defined as:
\begin{align}
f_{\text{sec}}(\text{QBER}) &= 1 - 2h(\text{QBER}) \\
h(p) &= -p\log_2 p - (1-p)\log_2(1-p)
\end{align}
where \( h(p) \) is the binary entropy function. As the QBER increases, the effective secure key rate drops sharply, potentially leading to a halt in key output if the error exceeds tolerable thresholds. To simulate the non-stationary behavior of \( G(t) \) under external disturbances such as storms or temperature surges, we introduce a first-order stochastic process to describe its temporal dynamics:
\begin{equation}
\dot{G}(t) = -\lambda_G \left(G(t) - \bar{G}\right) + \varepsilon_G(t)
\end{equation}
where \( \bar{G} \) is the nominal (steady-state) expected generation rate, \( \lambda_G > 0 \) is the regression coefficient characterizing the system's return speed to equilibrium, and \( \varepsilon_G(t) \) is a zero-mean white noise process representing high-frequency fluctuations caused by environmental uncertainties (e.g., optical misalignment or ground potential variations).

\subsection{Key Consumption Model $C_i(t)$}

Integrating QKD into power dispatch communication requires modeling how control tasks consume keys at different rates and security levels. Control messages fall into two main types: monitoring messages (e.g., PMU uploads, telemetry) with high frequency but low key demand, and execution messages (e.g., AGC, AVR, load shedding) that are security-critical and require prioritized key usage. Let $C_i(t)$ denote the instantaneous key consumption rate (in bits per second) for the $i$-th type of control task at time $t$, where $i = 1,2,\dots,T$. The key consumption can be expressed as:
\begin{equation}
C_i(t) = \alpha_i \cdot L_i \cdot \delta_i(t),
\end{equation}
where $L_i$ is the bit length of a single message for task type $i$, which can be determined based on relevant communication protocols such as IEC 60870-5-104 or IEC 61850. The parameter $\alpha_i$ represents the encryption strength coefficient. For standard symmetric encryption (e.g., AES-128), we define $\alpha_i = 1$, whereas for one-time pad (OTP) encryption, which requires a key bit for every message bit, we set $\alpha_i = L_i^{-1}$. The binary indicator $\delta_i(t)$ equals 1 if the $i$-th task is triggered at time $t$, and 0 otherwise. Considering that task activation is often stochastic—particularly for high-frequency controls such as AGC or AVR—we model the triggering process as a Poisson process $N_i(t)$ with mean activation rate $\lambda_i$. The key consumption during an infinitesimal time interval $dt$ can then be described as:
\begin{equation}
dN_i(t) \sim \mathrm{Poisson}(\lambda_i \, dt), \quad 
C_i(t) \cdot dt = \alpha_i \cdot L_i \cdot dN_i(t).
\end{equation}
In other words, within a small time interval $dt$, the probability that a task of type $i$ is triggered is approximately $\lambda_i dt$. If triggered, it consumes a fixed quantity of key material, equal to $\alpha_i L_i$.


\subsection{Key Inventory Dynamics and Safety Threshold Design}

Building on the models for key generation and consumption, we now formulate a dynamic model of the quantum key inventory that reflects real-time variations in the available cryptographic resources. This model serves not only to quantify the current state of the key pool but also supports control chain reconfiguration, dispatch strategy adjustments, and proactive system warnings under communication-constrained conditions.

Drawing an analogy to a reservoir system, we conceptualize the QKD key pool as a storage reservoir, where the key generation process $G(t)$ represents the inflow, and the total consumption rate $\sum_i C_i(t)$ across all control tasks acts as the outflow. Based on this analogy, we define the continuous-time dynamic model for the key inventory $K(t)$ as:
\begin{equation}
\dot{K}(t) = G(t) - \sum_{i=1}^{T} C_i(t)
\end{equation}
Here, $K(t)$ represents the total amount of usable key material in bits at time $t$. The function $G(t)$ denotes the instantaneous key generation rate from the QKD link, while $\sum_{i=1}^{T} C_i(t)$ captures the total instantaneous key consumption rate from all control tasks. Since the key pool cannot be stored indefinitely due to hardware constraints and cryptographic key freshness requirements, the above differential equation provides a realistic characterization of the key system’s operational behavior.

Under discrete-time implementation, which is typical in energy management systems (EMS) with sampling intervals such as $\Delta t = 100\,\mathrm{ms}$, we use the following difference equation:
\begin{equation}
K_{k+1} = K_k + \left( G_k - \sum_{i=1}^{T} C_{i,k} \right) \cdot \Delta t,
\end{equation}
where $K_k$ denotes the key inventory at the $k$-th sampling instant, and $G_k$, $C_{i,k}$ represent the sampled values of generation and consumption rates, respectively. This recursive formula can be used for real-time forecasting of future inventory levels, serving as a key availability status indicator for the dispatch module.

To ensure that critical control tasks in the power grid---such as primary frequency regulation, emergency load shedding, and control path reconfiguration---can still be executed even under key resource pressure, we introduce three inventory-related thresholds. These thresholds form the basis of a ``warn–reconfigure–protect'' tiered management mechanism for the key pool. The first is the \textit{safety threshold} $K_{\mathrm{safe}}$, defined as the minimum amount of key required to guarantee uninterrupted execution of the highest-priority task under extreme conditions (e.g., dispatch link failure or cyberattack). Assuming the task with the highest priority, denoted by index $i^*$, must operate reliably over a critical time window $\tau_{\mathrm{crit}}$, with triggering frequency $\lambda_{\max}$ and message length $L_{\max}$, the safety threshold is given by:
\begin{equation}
K_{\mathrm{safe}} = \alpha_{\max} \cdot L_{\max} \cdot \lambda_{\max} \cdot \tau_{\mathrm{crit}},
\end{equation}
where $\alpha_{\max}$ denotes the maximum key strength parameter used by the task, such as the one-time pad (OTP) encryption level.

The second is the \textit{reconfiguration threshold} $K_{\mathrm{th}}$, which triggers system-wide control chain reconfiguration actions such as task downgrading, dispatch prioritization, or redundant path activation. To ensure sufficient redundancy even after reconfiguration is triggered, the following condition must be satisfied:
\begin{equation}
K_{\mathrm{th}} > K_{\mathrm{safe}}.
\end{equation}

The third threshold is the \textit{capacity limit} $K_{\mathrm{cap}}$, which reflects the physical buffer size of the QKD hardware and the key lifecycle management constraints. When the inventory exceeds this value, surplus keys must be discarded to ensure freshness and compliance with cryptographic expiration policies.

\subsection{Prediction and Uncertainty Estimation}

Due to uncertainty in quantum key generation and consumption, control centers must forecast key inventory $K(t)$ to ensure secure task execution. We propose a hybrid LSTM–Kalman filter model that combines LSTM's nonlinear prediction strength with Kalman filtering’s real-time error correction, enabling robust short-term forecasting with confidence intervals for QKD-enhanced SCADA systems. We begin by defining the system observation vector at time step $k$:
\begin{equation}
\mathbf{y}_k = \left[G_k,\; C_{1,k},\; C_{2,k},\dots,\; C_{T,k}\right]^\top
\end{equation}
where $G_k$ is the observed key generation rate at time step $k$, and $C_{i,k}$ is the key consumption rate associated with the $i$-th control task. These variables can be acquired in real time from the QKD link monitoring module and SCADA message tracking system. Using historical time-series data, we train a multivariate LSTM model to produce one-step-ahead predictions of the key system dynamics:
\begin{align}
\Bigl[
\,\hat{G}_{k|k-1},\; & \hat{C}_{1,k|k-1},\; \hat{C}_{2,k|k-1},\; \dots,\; \hat{C}_{T,k|k-1} 
\,\Bigr]^\top \nonumber \\
&= \mathrm{LSTM}(\mathbf{y}_{k-1},\; \mathbf{y}_{k-2},\; \dots,\; \mathbf{y}_{k-w})
\end{align}
where $w$ is the sliding window size. The outputs $\hat{G}_{k|k-1}$ and $\hat{C}_{i,k|k-1}$ represent the predicted values of generation and task-specific consumption rates, respectively.

To correct for systemic model errors and to quantify the uncertainty in state propagation, we apply an Extended Kalman Filter (EKF) to update the key system state vector and estimate the associated state covariance. The system state vector is defined as:
\begin{equation}
\mathbf{x}_k = \left[\hat{G}_k,\; \hat{K}_k \right]^\top, \quad
\hat{\mathbf{x}}_{k|k-1} = f(\hat{\mathbf{x}}_{k-1}),
\end{equation}
where $\hat{K}_k$ is the predicted key inventory at time $k$. The predicted state is updated using the standard Kalman correction equation:
\begin{equation}
\hat{\mathbf{x}}_k = \hat{\mathbf{x}}_{k|k-1} + \mathbf{K}_k \left( \mathbf{y}_k - \mathbf{H} \hat{\mathbf{x}}_{k|k-1} \right),
\end{equation}
in which $\mathbf{K}_k$ is the Kalman gain and $\mathbf{H}$ is the observation matrix. The process model $f(\cdot)$ is based on the discrete-time inventory update rule derived from earlier sections:
\begin{equation}
\hat{K}_k = \hat{K}_{k-1} + \left( \hat{G}_k - \sum_{i=1}^T \hat{C}_{i,k} \right) \cdot \Delta t.
\end{equation}
From the corrected state and its error covariance, we obtain a Gaussian approximation of the inventory distribution. For example, the 95\% confidence interval for the key inventory at time step $k$ is given by:
\begin{equation}
K_k \sim \mathcal{N}(\hat{K}_k,\; P_{K,k}), \quad
\mathrm{CI}_{95\%} = \hat{K}_k \pm 1.96 \sqrt{P_{K,k}},
\end{equation}
where $P_{K,k}$ is the variance of the inventory estimate. This interval provides a statistically grounded boundary on key depletion risk and can be used to trigger early warnings or remedial actions, such as link switching, task degradation, or backup path activation.

The entire prediction process is computationally efficient. With a state dimension of $n_x = 2$, both LSTM inference and Kalman update can be executed within 50 milliseconds, enabling online deployment in modern EMS platforms.

\subsection{Dynamic Key Control Workflow and Performance Metrics}

Building upon the preceding models for key generation, consumption, prediction, and uncertainty quantification, this section presents a practical \textit{dynamic key control workflow} for QKD-enhanced dispatch systems. It also introduces a multi-dimensional set of \textit{performance evaluation metrics}, designed to quantify the impact of key resource fluctuations on control task execution, system stability, and dispatch responsiveness.

During each sampling interval $k$ (e.g., $\Delta t = 100$ ms), the control center can follow the steps below to monitor and update the quantum key inventory in real time and trigger responsive control strategies:

To systematically evaluate the influence of key availability on power system control behavior, we further define the following set of performance metrics:

\textbf{1) Task Success Rate ($P_{\mathrm{succ}}$):}  
This metric measures the proportion of control tasks that were encrypted and executed without delay or downgrade during the dispatch cycle:
\begin{equation}
P_{\mathrm{succ}} = \frac{N_{\mathrm{success}}}{N_{\mathrm{trigger}}}
\end{equation}
where $N_{\mathrm{trigger}}$ is the total number of control tasks triggered, and $N_{\mathrm{success}}$ is the number of those successfully executed with appropriate key security. This reflects the coverage capability of different key scheduling strategies.

\textbf{2) Maximum Frequency Deviation ($\Delta f_{\max}$):}  
Under frequency control scenarios (e.g., AGC), insufficient key resources may delay or block commands, causing transients in grid frequency. The worst-case deviation during the observation period is defined as:
\begin{equation}
\Delta f_{\max} = \max_{t\in [t_0, t_1]} |\Delta f(t)|
\end{equation}
This metric reflects the control reliability degradation caused by communication or cryptographic limitations.

\textbf{3) Key Utilization Rate ($\eta$):}  
This metric quantifies how efficiently the generated keys are utilized across control tasks. It is defined as:
\begin{equation}
\eta = \frac{\sum_{k} \sum_i C_{i,k}}{\sum_k G_k}
\end{equation}
Higher utilization indicates better alignment between supply and demand; low utilization may reflect redundancy or inefficient dispatch strategies.

\textbf{4) Time to Resilience Recovery (TRR):}  
This metric captures the duration required for the system to return to stable operation following a key-induced control failure or deviation. It is defined as:
\begin{equation}
\mathrm{TRR} = \int_{t_{\text{fault}}}^{t_{\text{recov}}} \mathbf{1}_{|\Delta f(t)| > \Delta f_{\text{safe}}} \, dt
\end{equation}
where $\Delta f_{\text{safe}}$ is the maximum acceptable frequency deviation. This indicator evaluates the resilience of the system under key resource stress.

\section{Control Chain Reconfiguration}

\subsection{Reconfiguration Modeling}

To address the risk of critical control chain failure caused by dynamic fluctuations in quantum key availability, this section proposes a multi-layer network modeling framework that captures the coupling between communication, control, and physical layers. A quantitative objective function is further formulated to balance key consumption, control effectiveness, and system resilience, forming the foundation for subsequent optimization. The QKD-enabled SCADA system is abstracted as a directed multilayer graph:
\begin{equation}
\mathcal{G} = (\mathcal{V}^{\mathrm{com}},\;\mathcal{V}^{\mathrm{ctr}},\;\mathcal{V}^{\mathrm{phy}};\;
               \mathcal{E}^{\mathrm{com}},\;\mathcal{E}^{\mathrm{ctr}},\;\mathcal{E}^{\mathrm{phy}})
\end{equation}
Here, $\mathcal{V}^{\mathrm{com}}$ represents communication nodes (e.g., QKD terminals, switches), $\mathcal{V}^{\mathrm{ctr}}$ denotes control task nodes (e.g., AGC, AVR, protection signals), and $\mathcal{V}^{\mathrm{phy}}$ includes physical equipment such as buses, generators, and loads. The edge sets $\mathcal{E}^{(\cdot)}$ capture both intra-layer and inter-layer interactions. Each control chain $\ell$ is defined as a directed path $\mathcal{P}_\ell \subseteq \mathcal{G}$ that originates from the control center, passes through communication nodes, and actuates physical components. It is characterized by a feature vector:
\begin{equation}
\boldsymbol{\phi}_\ell = \left(
d_\ell,\;
\tau_\ell,\;
\beta_\ell,\;
\gamma_\ell
\right)
\end{equation}
where $d_\ell$ denotes the key resource required per unit time, $\tau_\ell$ is the end-to-end latency tolerance, $\beta_\ell$ reflects the control priority or impact (e.g., frequency response weight), and $\gamma_\ell$ captures the reconfiguration cost associated with activating or switching the chain.

To adapt to time-varying key inventory and ensure continuous execution of prioritized control tasks, the system must dynamically decide the state of each control chain. Let $x_{\ell,k} \in \{0,1,2\}$ denote the state of control chain $\ell$ at discrete time step $k$, where:
\[
x_{\ell,k}=
\begin{cases}
0, & \text{chain deactivated} \\
1, & \text{chain active with downgraded encryption} \\
2, & \text{chain active with full encryption (e.g., OTP)}
\end{cases}
\]
Then, the total key consumption at time $k$ is given by:
\begin{equation}
C_k = \sum_{\ell\in\mathcal{L}}\;
      d_\ell\left( \frac{\alpha_{\mathrm{AES}}}{\Delta t} \cdot \mathbf{1}_{x_{\ell,k}=1}
                 + \frac{\alpha_{\mathrm{OTP}}}{\Delta t} \cdot \mathbf{1}_{x_{\ell,k}=2} \right)
\end{equation}
where $\alpha_{\mathrm{AES}}$ and $\alpha_{\mathrm{OTP}}$ denote the key usage coefficients under AES and OTP encryption, respectively, and $\Delta t$ is the control dispatch interval. $\mathbf{1}_{(\cdot)}$ is the indicator function.

To prevent abrupt key depletion and ensure reliability of mission-critical chains, the decision framework includes a chance-constrained key inventory requirement:
\begin{equation}
\Pr\left( C_k \le \hat{K}_k - \eta_{\mathrm{buf}} \right) \ge 1 - \varepsilon
\end{equation}
Here, $\hat{K}_k$ denotes the predicted key inventory at time $k$, $\eta_{\mathrm{buf}}$ is the safety buffer margin, and $\varepsilon$ is the risk tolerance (e.g., 0.05 corresponding to 95\% confidence). This constraint ensures that the key allocation decisions remain feasible under prediction uncertainty.

Based on the above structure, we construct the overall control loss function $\mathcal{L}_k$ to evaluate performance at each dispatch step. The function integrates power system stability deviation, task execution penalties, and reconfiguration overhead:
\begin{equation}
\begin{aligned}
\mathcal{L}_k =&
\underbrace{ \sum_{n\in\mathcal{N}} w_f \left|\Delta f_{n,k}\right|
           + \sum_{m\in\mathcal{M}} w_v \left|\Delta V_{m,k}\right| }_{\text{grid stability deviation}} \\
&+ \underbrace{ \sum_{\ell} \left( \xi_{\mathrm{drop}}\,\mathbf{1}_{x_{\ell,k}=0}
           + \xi_{\mathrm{deg}}\,\mathbf{1}_{x_{\ell,k}=1} \right) }_{\text{task degradation and failure cost}} \\
&+ \underbrace{ \sum_{\ell} \gamma_\ell \cdot \mathbf{1}_{x_{\ell,k} \neq x_{\ell,k-1}} }_{\text{chain reconfiguration cost}}
\end{aligned}
\end{equation}
In this formulation, $\Delta f_{n,k}$ and $\Delta V_{m,k}$ denote the frequency and voltage deviations at bus $n$ and node $m$ at time $k$, while $w_f$ and $w_v$ are corresponding weight factors. The terms $\xi_{\mathrm{drop}}$ and $\xi_{\mathrm{deg}}$ quantify the cost of fully dropped or downgraded control tasks, and $\gamma_\ell$ penalizes frequent switching of chain $\ell$ to discourage excessive reconfiguration. This loss function encapsulates the essential trade-off between resource-aware encryption scheduling and power system control effectiveness. Compared to static chain architectures or simple task-priority-based schemes, the proposed model enables real-time task-level adaptation to key availability, maintaining high control success rates and continuous secure operation in dynamically constrained environments.

\subsection{Temporal Optimization Formulation}

To enable rolling adjustments of control chain status and dynamic scheduling of quantum key resources, we formulate a temporal optimization model that operates over a fixed control interval $\Delta t$. The model aims to minimize overall control performance loss over a predictive horizon while considering forecast uncertainty in key inventory, power system dynamics, and the cost of switching control chains. This results in a practically solvable mixed-integer chance-constrained optimization framework.

Let the rolling prediction horizon be $H$ steps. At each dispatch step $k$, the model optimizes the activation sequence $\{x_{\ell,t}\}_{t=k}^{k+H-1}$ for each control chain $\ell$ across the window. The objective function is defined as:
\begin{equation}
\min_{\{x_{\ell,t}\}} \quad
      \sum_{t=k}^{k+H-1} \left( \mathcal{L}_t + \lambda \left\| \boldsymbol{\Delta f}_t \right\|_2^2 \right)
\end{equation}
Here, $\mathcal{L}_t$ denotes the control chain loss function at time $t$, defined in Equation (8), while $\boldsymbol{\Delta f}_t$ represents the vector of frequency deviations at time $t$, and $\lambda$ is a weight parameter balancing control quality and resource usage.

The optimization is subject to three categories of constraints:

\textbf{(1) Key Resource Chance Constraint:}  
For each $t \in [k, k+H-1]$, the cumulative key consumption $C_t$ arising from selected control chains must satisfy a probabilistic lower bound on the available key inventory:
\begin{equation}
\Pr\left( C_t \le \hat{K}_t - \eta_{\mathrm{buf}} \right) \ge 1 - \varepsilon
\end{equation}
Here, $\hat{K}_t$ is the forecasted key inventory at time $t$, $\eta_{\mathrm{buf}}$ denotes a safety buffer, and $\varepsilon$ is the risk tolerance level (e.g., 0.05 for 95\% confidence). This chance constraint ensures that control actions remain feasible under uncertainty and can be transformed into a deterministic second-order cone (SOC) constraint using techniques such as Chebyshev inequality or Boole–Bonferroni approximation.

\textbf{(2) Control Chain State Constraints:}  
Each control chain can only assume one of three discrete states at any time:
\begin{equation}
x_{\ell,t} \in \{0, 1, 2\}, \quad \forall \ell, \; \forall t \in [k, k+H-1]
\end{equation}
Specifically, $x_{\ell,t}=0$ indicates chain deactivation, $x_{\ell,t}=1$ implies downgraded encryption (e.g., AES), and $x_{\ell,t}=2$ represents full-strength encryption (e.g., OTP).

\textbf{(3) Power System Dynamics Constraint (State-Space Form):}  
To model how control decisions affect system-level dynamics such as frequency and voltage trajectories, we adopt a linearized discrete-time state-space model:
\begin{equation}
\mathbf{x}_{\mathrm{sys}}(t+1) = \mathbf{A} \mathbf{x}_{\mathrm{sys}}(t) + \mathbf{B} \mathbf{u}(t)
\end{equation}
In this expression, $\mathbf{x}_{\mathrm{sys}}(t)$ denotes the system state vector (e.g., frequency deviations, voltage angles), and $\mathbf{u}(t)$ is the control input vector. The matrices $\mathbf{A}$ and $\mathbf{B}$ define the system dynamics, obtained through linearization of the power system around the operating point. The control input $\mathbf{u}(t)$ is mapped from the chain activation status using a constraint of the form:
\begin{equation}
\mathbf{u}(t) = \mathbf{U}(x_{\ell,t})
\end{equation}
This mapping ensures logical consistency between the activated chains and the actual control commands dispatched, such as AGC setpoints or load shedding signals.

The resulting optimization problem is a Mixed-Integer Second-Order Cone Programming (MISOCP) model due to the presence of integer variables $x_{\ell,t}$, nonlinear objectives, and SOC constraints arising from uncertainty. To ensure tractability and scalability, for small- to medium-scale systems, commercial solvers such as CPLEX or Gurobi can be used directly with branch-and-bound methods. For larger systems, solution speed can be enhanced via decomposition techniques such as column generation, Benders decomposition, or penalty-based relaxation heuristics.

\section{Stackelberg Cooperative Game and MPEC-Based Solution}

In QKD-encrypted power systems, quantum key resources are shared between TSO and DSO, creating inherent coordination conflicts. While the TSO focuses on frequency support and grid stability, the DSO prioritizes local voltage and load control. This interaction is modeled as a bi-level Stackelberg game, with the TSO as leader and the DSO as follower. Let $\mathbf{x}^{\mathrm{T}}$ denote the TSO’s decision variables over controlled secure chains (e.g., activation states and encryption modes), and $\mathbf{x}^{\mathrm{D}}$ represent the corresponding DSO decisions. The bi-level problem is then defined as:
\begin{align}
\min_{\mathbf{x}^{\mathrm{T}}} \quad & 
    J^{\mathrm{T}}(\mathbf{x}^{\mathrm{T}}, \mathbf{x}^{\mathrm{D}}) = 
    \sum_{t=k}^{k+H-1} \left[ \mathcal{L}^{\mathrm{T}}_t + 
    \lambda^{\mathrm{T}} \|\Delta \mathbf{f}^{\mathrm{T}}_t\|_2^2 \right] \nonumber \\
\text{s.t.} \quad & 
    \mathbf{x}^{\mathrm{T}} \in \Omega^{\mathrm{T}}, \nonumber \\
& 
    \mathbf{x}^{\mathrm{D}} \in 
    \arg\min_{\mathbf{x}} \left\{
    \sum_{t=k}^{k+H-1} \left[ \mathcal{L}^{\mathrm{D}}_t + 
    \lambda^{\mathrm{D}} \|\Delta \mathbf{f}^{\mathrm{D}}_t\|_2^2 \right]
    \right. \nonumber \\
& \qquad \left. \middle| \; \mathbf{x} \in \Omega^{\mathrm{D}}(\mathbf{x}^{\mathrm{T}}) \right\}
\end{align}
Here, $\mathcal{L}^{\mathrm{T}}_t$ and $\mathcal{L}^{\mathrm{D}}_t$ denote the control chain losses for TSO and DSO, respectively, incorporating frequency deviation penalties, control errors, and encryption switching costs. The feasible sets $\Omega^{\mathrm{T}}$ and $\Omega^{\mathrm{D}}$ include power system dynamics, key budget constraints, and discrete mode restrictions. The coupling is induced by a shared probabilistic quantum key budget constraint:
\begin{equation}
\Pr \left( C^{\mathrm{T}}_t + C^{\mathrm{D}}_t \leq \hat{K}_t - \eta_{\text{buf}} \right) \geq 1 - \varepsilon, \quad \forall t
\end{equation}

To reformulate the bi-level problem into a numerically tractable single-level model, we derive the Karush-Kuhn-Tucker (KKT) conditions for the DSO's lower-level optimization. Let the DSO's constraints be expressed as:
Nonlinear inequalities: $\mathbf{g}(\mathbf{x}^{\mathrm{T}}, \mathbf{x}^{\mathrm{D}}) \le 0$; Second-order cone (SOC) constraints: $\mathbf{h}(\mathbf{x}^{\mathrm{T}}, \mathbf{x}^{\mathrm{D}}) \in \mathcal{K}$; Integer variable relaxations: $x_{\ell,t} \in \{0,1,2\}$. Introducing Lagrange multipliers $\boldsymbol{\lambda}$ and $\boldsymbol{\mu}$ for $\mathbf{g}$ and $\mathbf{h}$ respectively, the KKT system becomes:
\begin{align}
\nabla_{\mathbf{x}^{\mathrm{D}}} J^{\mathrm{D}} 
+ \boldsymbol{\lambda}^\top \nabla_{\mathbf{x}^{\mathrm{D}}} \mathbf{g}
+ \boldsymbol{\mu}^\top \nabla_{\mathbf{x}^{\mathrm{D}}} \mathbf{h} = 0 \\
\lambda_i \cdot g_i = 0, \quad \forall i; \quad 
\mathbf{h} \in \mathcal{K},\; \boldsymbol{\mu} \in \mathcal{K}^*,\;
\langle \mathbf{h}, \boldsymbol{\mu} \rangle = 0 \\
x_{\ell,t}(1 - x_{\ell,t}) = 0, \quad \forall \ell, t
\end{align}
Embedding the above complementarity system into the upper-level constraints yields a Mathematical Program with Equilibrium Constraints (MPEC), specifically a Mixed-Integer Program with Complementarity Constraints (MPCC).

To solve the resulting MPCC efficiently, we propose a Level Decomposition with Complementarity Pruning (LD-CP) algorithm that leverages the sparse coupling structure between the TSO (leader) and DSO (follower), as shown in Algorithm 2. Specifically, the original bi-level formulation 
$\min_{\mathbf{x}^{\mathrm{T}}} J^{\mathrm{T}}(\mathbf{x}^{\mathrm{T}}, \mathbf{x}^{\mathrm{D}})$ 
with $\mathbf{x}^{\mathrm{D}} \in \arg\min J^{\mathrm{D}}(\mathbf{x}^{\mathrm{T}}, \cdot)$ 
is decoupled into alternating rounds of leader–follower interaction. At each iteration $p$, the leader’s decision $\mathbf{x}^{\mathrm{T},(p)}$ is fixed, and the follower responds by solving its own optimization subproblem. From the resulting KKT system, we extract the active complementarity set 
$\mathcal{C}^{(p)} \subseteq \{ (i,j) \mid \lambda_i \cdot g_i = 0,\ \langle h_j,\mu_j \rangle = 0 \}$, 
which reflects the subset of dual constraints that are exactly satisfied under current conditions. These active pairs are then encoded back into the upper-level problem via binary indicator variables $z_i \in \{0,1\}$ that govern whether $g_i \le 0$ or $\lambda_i \le 0$ holds at iteration $p$. The resulting upper-level program introduces a proximal regularization term 
$\rho \|\mathbf{x}^{\mathrm{T}} - \mathbf{x}^{\mathrm{T},(p)}\|^2$ 
to stabilize updates and improve convergence. After solving for a new leader decision $\mathbf{x}^{\mathrm{T},(p+1)}$, the active set is pruned by removing invalid complementarity pairs based on feasibility and duality conditions, yielding a refined $\mathcal{C}^{(p+1)}$. This iterative process continues until the update norm satisfies 
$\|\mathbf{x}^{\mathrm{T},(p+1)} - \mathbf{x}^{\mathrm{T},(p)}\| < \delta$, 
or a predefined maximum iteration count is reached.

\section{Simulation Platform Implementation}

To enable comprehensive modeling and verification of QKD-enhanced SCADA control chains in power systems, we develop an end-to-end simulation platform consisting of five tightly coupled modules: link disturbance injection, key generation and management, protocol stack binding, control chain configuration, and visualization monitoring. The platform adopts a modular design, supports deployment on standalone or clustered environments, and is compatible with hardware-in-the-loop (HIL) setups such as OPAL-RT or RTDS. 

\subsection{Optical Attenuation and Link Disruption Injection (OpenQKD-Sim)}

We enhance the OpenQKD-Sim module to inject realistic non-weather-related disturbances at the physical layer. The instantaneous optical attenuation of the link is modeled as:
\begin{equation}
\eta(t) = 10^{-\left( \alpha_0 + \Delta\alpha(t) \right) L / 10}
\end{equation}
where $\alpha_0$ denotes the nominal attenuation coefficient (in dB/km), $L$ is the fiber length, and $\Delta\alpha(t) \sim \mathcal{N}(0, \sigma_\alpha^2)$ captures random fluctuations due to physical factors such as thermal drift and fiber vibration. Link breakage events are modeled as a Poisson process:
\begin{equation}
\mathrm{d}N(t) \sim \mathrm{Poisson}(\lambda_{\text{break}}\,\mathrm{d}t)
\end{equation}
with $\lambda_{\text{break}}$ denoting the expected failure frequency. Scenario parameters such as attenuation bounds, break intervals, and restoration latency are defined in YAML, with real-time output of the key generation rate $G(t)$.

\subsection{QKD Key Generation and Management API}

A Python-based gRPC interface implements the QKD key pool management layer. The \texttt{KeyServer} module provides two key methods:
\begin{align}
\text{GetKey}(n) &\rightarrow \{k_1, \dots, k_n\} \\
\text{KeyPoolStatus}() &\rightarrow K(t)
\end{align}
where $K(t)$ denotes the real-time size of the key inventory. The \texttt{KeyClient}, embedded in the SCADA control interface, uses asynchronous pulling with local buffering to decouple key fetch latency from critical control paths. Key dynamics follow the differential model proposed in Section 4.4, and state updates are published to the control optimizer via Redis Pub/Sub.

\subsection{Protocol Stack Binding (Q3P + UDP + IEC 60870-5-104)}

We integrate the Q3P protocol with UDP/IP and IEC 104 to construct a secure, backward-compatible stack. The transmission layer is structured as:
\[
\text{UDP/IP} \rightarrow \text{Q3P (Key Index + MAC)} \rightarrow \text{IEC 104 ASDU}
\]
The Q3P module embeds a 16-byte authentication tag and one-time-pad (OTP) index into the IEC 104 payload without changing its frame structure. Each message $m$ requires a key consumption of:
\begin{equation}
d(m) = \alpha(m) \cdot |m|, \quad \alpha(m) =
\begin{cases}
1, & \text{AES-128} \\
|m|^{-1}, & \text{OTP}
\end{cases}
\end{equation}
Transmission latency $\tau_{\text{link}}(t)$ is dynamically determined by OpenQKD-Sim and injected via controlled delay calls (e.g., \texttt{sleep} operations) to simulate effects such as fiber bending or device buffering.

\subsection{Control Chain Configuration and Visualization Monitoring}

Control chains are defined using a Python domain-specific language (DSL), e.g.:
\texttt{Chain(id, task\_type, priority, crypto\_mode, path=[node1, node2, ...])}. These configurations are uploaded via REST API and parsed by the scheduler. The scheduling engine monitors the real-time key inventory $K(t)$, fiber availability $\eta(t)$, and control deviations $\Delta f(t)$ to compute the optimal activation vector $\mathbf{x}_t$, as defined in the optimization model of Section 5. A Grafana dashboard, backed by InfluxDB, provides real-time visualization of $\hat{K}_t$, task capacity $C_t$, node deviations $\Delta f_{n,t}$, and link-level events such as breakage or recovery. If thresholds like $K_t < K_{\text{safe}}$ or $\Delta f_{n,t} > \Delta f_{\max}$ are violated, alert notifications are triggered via Slack or email.

\section{Case Study}
\subsection{Experiment Setup}
This study presents a comprehensive and repeatable simulation framework to evaluate the proposed QKD-enhanced SCADA control chain reconfiguration strategy. The IEEE 39-bus and 118-bus systems are selected as benchmark testbeds. The 39-bus system serves as a medium-scale case for verifying system stability and efficiency, while the 118-bus system is partitioned into one transmission zone and three distribution areas to support multi-agent coordination scenarios. The SCADA architecture follows a four-layer topology—Control Center, Gateway, RTU, and IED—with IEC 60870-5-104 communication protocol integrated with Q3P encryption. Optical link disturbances are injected via OpenQKD-Sim, assuming fiber lengths of 20 km (39-bus) and 50 km (118-bus), with baseline attenuation of 0.2 dB/km and Gaussian variations (0.04 dB/km). Link failures are modeled as Poisson-distributed events (mean rate: one per 5000s), lasting 3–5 seconds uniformly sampled. Five experimental scenarios are constructed: S-1 (static chain + static keys), S-2 (static chain + dynamic keys), S-3 (reconfigurable chain + dynamic keys, single-agent), S-4 (cooperative transmission-distribution control without game-theoretic optimization), and S-5 (Stackelberg-based multi-agent reconfiguration with MPEC and LD-CP algorithm). In S-5, a bi-level optimization is performed every 100 ms, where the TSO’s decision is updated using the LD-CP method based on complementarity set pruning and proximal regularization. Each experiment is repeated over 30 Monte Carlo runs to ensure statistical robustness. 

\subsection{IEEE 39-bus Case}

In this subsection, we conduct a comprehensive quantitative evaluation of the proposed method (S-3) against two baseline scenarios—static chain with static key (S-1), and static chain with dynamic key (S-2). 

\begin{table}[htbp]
\centering
\small
\caption{Performance comparison across scenarios}
\begin{tabular}{lccc}
\toprule
\textbf{Metric} & \textbf{S-1} & \textbf{S-2} & \textbf{S-3} \\
\midrule
Success Rate $P_{\text{succ}}$ & 0.72 & 0.88 & 0.97 \\
Max Frequency Deviation $\Delta f_{\max}$ (Hz) & 0.26 & 0.17 & 0.08 \\
Key Utilization $\eta$ & 0.30 & 0.57 & 0.83 \\
\bottomrule
\end{tabular}
\label{tab:compact-metrics}
\end{table}

\begin{figure}[!t]
\centering
\includegraphics[width=\columnwidth]{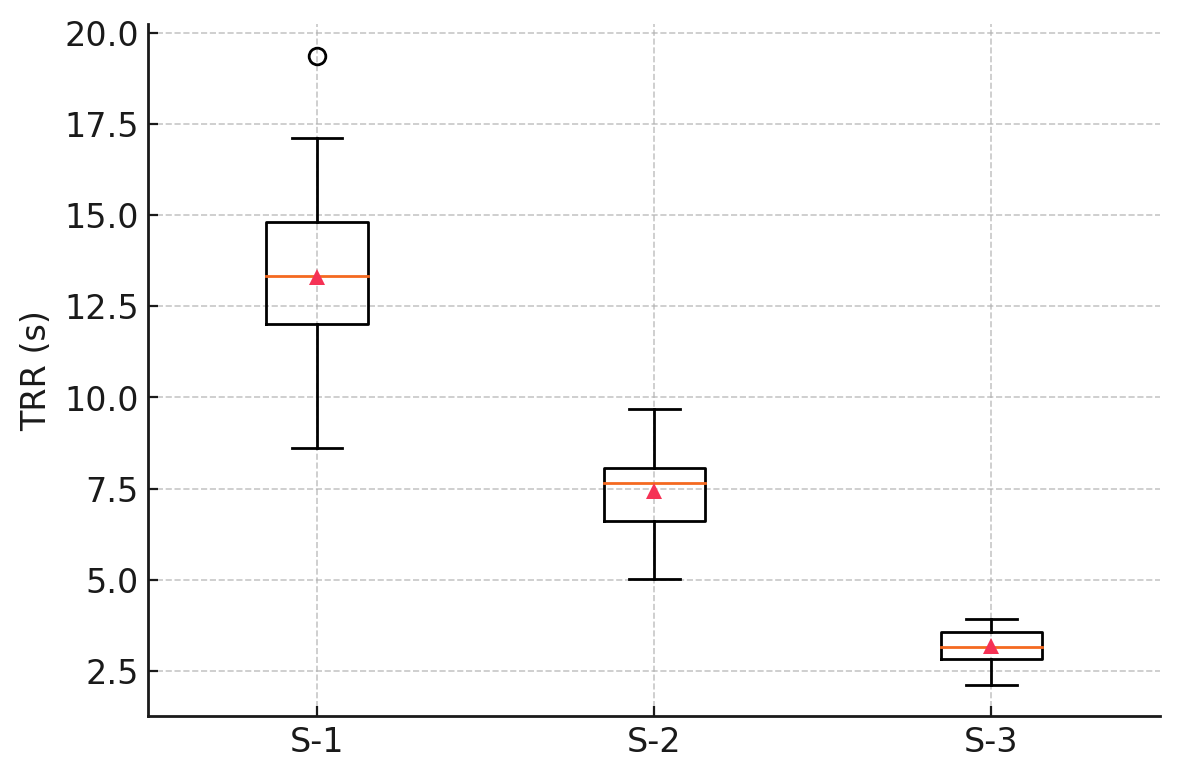}
\caption{Control Recovery Time Distribution}
\label{fig_sim}
\end{figure}

As shown in the aggregated metrics, S-1 exhibits the weakest overall performance: the task success rate remains at just 72\%, the frequency deviation peaks at 0.26~Hz, and the key utilization rate is below 30\%. This reflects the system’s high vulnerability to key exhaustion and communication disruptions under static configurations. When dynamic key generation is introduced in S-2, the task success rate increases to 88\%, and $\Delta f_{\max}$ drops to 0.17~Hz, indicating that continuous key supply significantly mitigates the risk of depletion. However, the key utilization rate only improves to 57\%, suggesting inefficiencies due to persistent link rigidity. In contrast, S-3 combines dynamic key management with adaptive link reconfiguration and task reprioritization, yielding the highest task success rate (97\%), the smallest frequency deviation (0.08~Hz), and the most efficient key utilization (83\%). These improvements validate the synergistic benefits of proactive coordination in both communication and control layers.

The 30-run TRR distributions further support these findings. S-1 demonstrates prolonged recovery times mostly in the 12–16~s range, with occasional outliers extending to 18–19~s, highlighting its reliance on passive recovery mechanisms. S-2 exhibits a shorter and tighter distribution between 6–9~s, though occasional delays still occur due to fixed communication topology. In sharp contrast, S-3 displays a highly compact TRR distribution centered around 2.5–3.5~s, with negligible variance and no outliers. This indicates that the integration of key-aware control rescheduling and real-time link reconfiguration enables rapid mitigation of control failures, minimizing frequency violations and downtime. Taken together, the results confirm that the proposed S-3 approach achieves superior resilience and resource efficiency under realistic operational constraints.

\subsection{Fairness Metrics under Stackelberg Coordination}

In this subsection, we critically analyse the fairness performance obtained in the Stackelberg scenarios by interpreting three complementary visualisations: a violin–box hybrid plot of the fairness index $F_K$, a colour-coded scatter plot relating LD-CP iteration count to $F_K$, and a heat-map that maps the standard deviation of $F_K$ over the joint space of iteration budget and link-break frequency.

\begin{figure}[!t]
\centering
\includegraphics[width=\columnwidth]{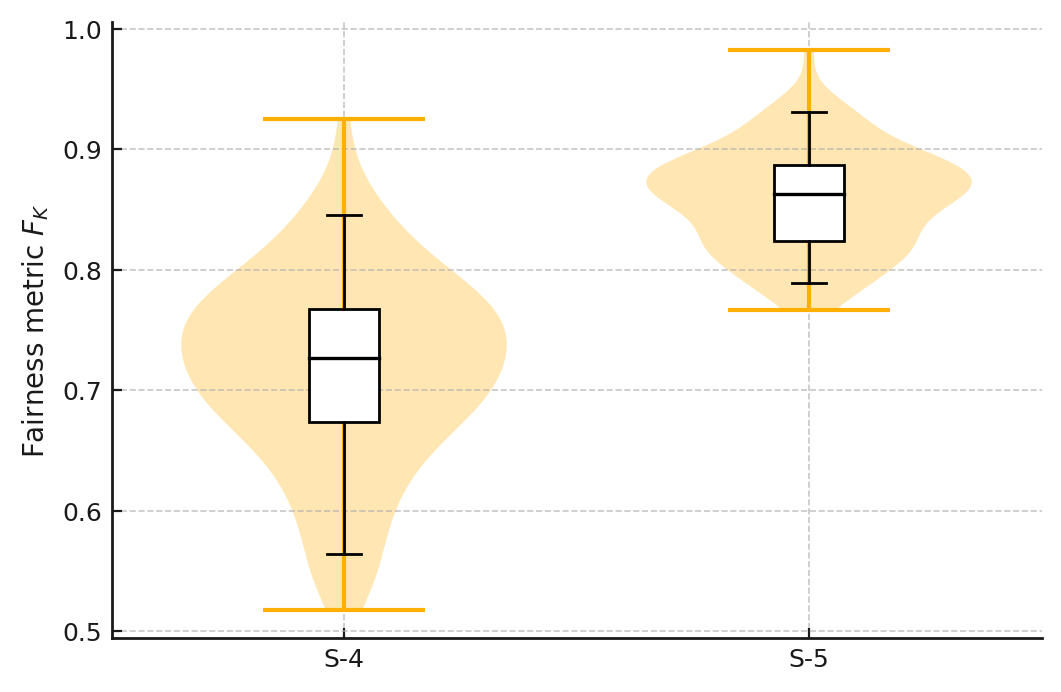}
\caption{Violin–Box Plot of Key-Fairness Distribution}
\label{fig_sim}
\end{figure}

\begin{figure}[!t]
\centering
\includegraphics[width=\columnwidth]{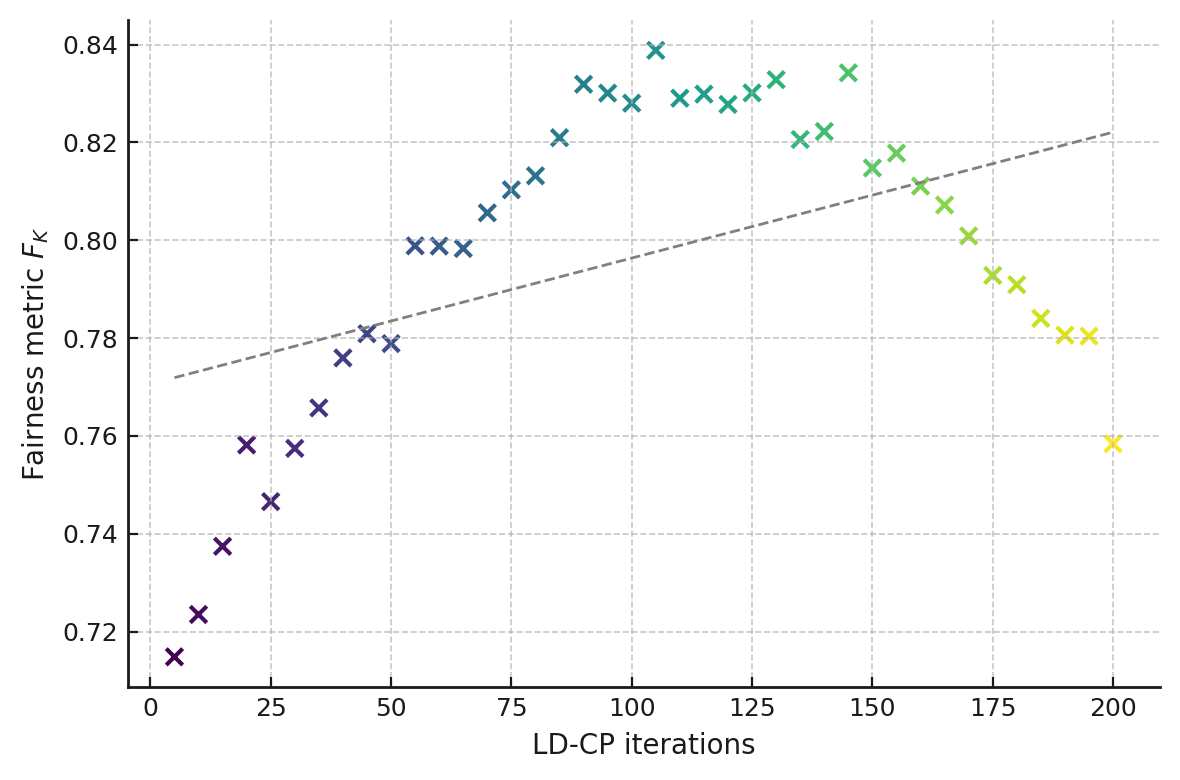}
\caption{LD-CP Iteration Count versus Achieved Fairness}
\label{fig_sim}
\end{figure}

\begin{figure}[!t]
\centering
\includegraphics[width=\columnwidth]{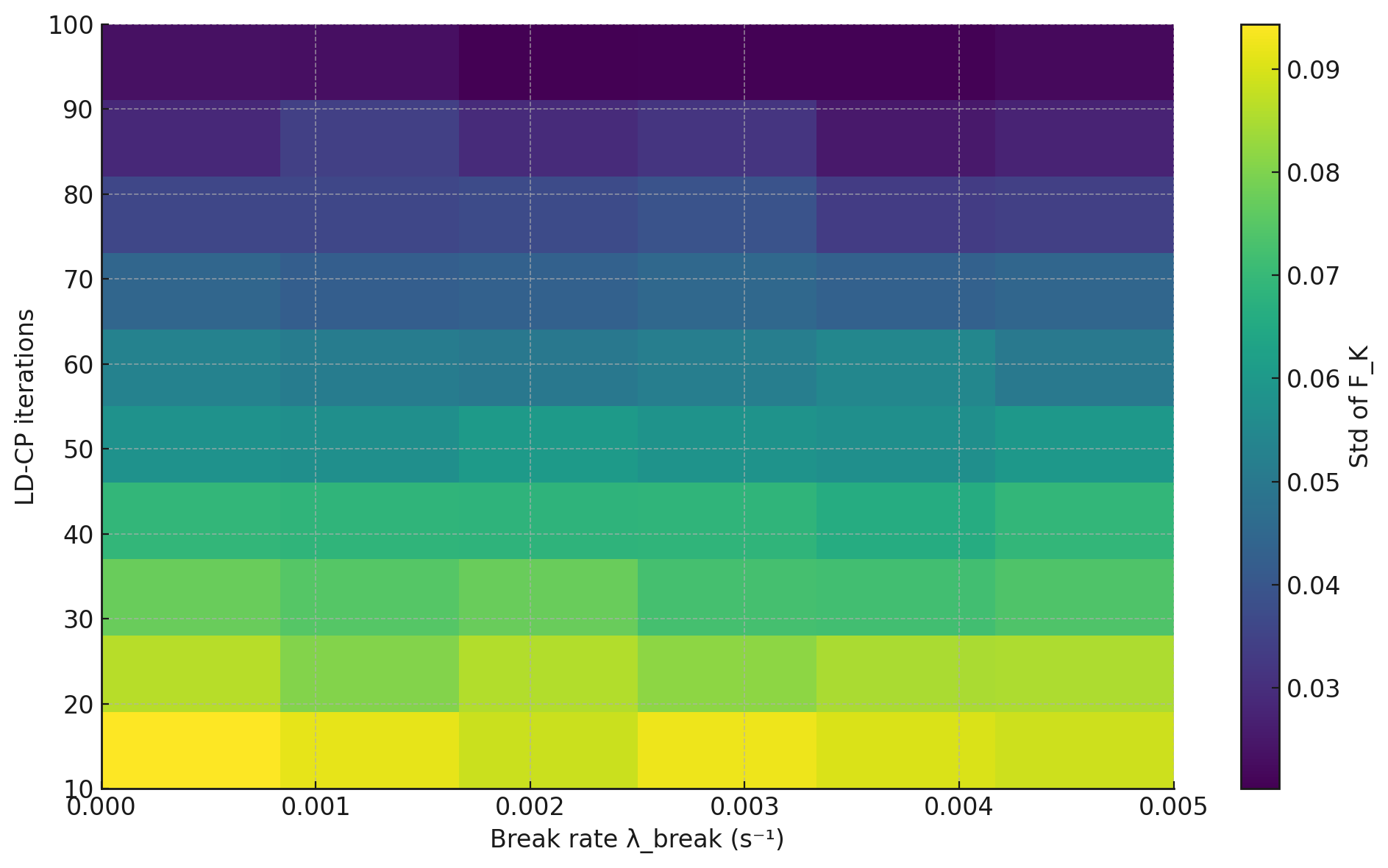}
\caption{Heat-Map of Fairness Variability as a Function of Iteration Budget and Link-Break Frequency}
\label{fig_sim}
\end{figure}

The violin–box plot reveals a pronounced improvement when the LD-CP-driven Stackelberg mechanism (S-5) replaces the static co-operative baseline (S-4).  Whereas S-4 centres around a median fairness of approximately $0.72$ with a visibly skewed tail dipping below $0.65$, S-5 shifts the entire distribution upwards to a median near $0.86$ and compresses its inter-quartile range by roughly half.  Because $F_K$ is a Jain-style index, this uplift signifies that the TSO and DSO obtain a far more balanced share of the finite quantum-key pool while run-to-run variability is simultaneously suppressed.  The plot therefore confirms that LD-CP’s complementarity pruning not only raises expected equity but also mitigates extreme imbalance events.

The iteration–fairness scatter plot complements this insight by quantifying the efficiency–fairness trade-off.  Fairness ascends steeply from about $0.70$ at five iterations to a plateau around $0.83$–$0.84$ once the iteration budget reaches $80$–$100$, after which marginal gains vanish.  The dashed regression line illustrates the overall positive slope, yet the curved cloud of points exposes diminishing returns: the interval between $25$ and $60$ iterations delivers the bulk of the improvement, whereas pushing beyond ${\sim}120$ iterations contributes little besides computational overhead.  Operationally, this suggests that capping the LD-CP loop at $50$–$80$ iterations provides an attractive balance between convergence speed and final equity.

Finally, the heat-map projects the standard deviation of $F_K$ onto the plane spanned by link-break frequency $\lambda_{\text{break}}$ and iteration count $N_{\text{iter}}$.  Three patterns emerge.  First, variance rises almost linearly with disturbance rate, confirming that heavier communication stress naturally degrades fairness.  Second, increasing $N_{\text{iter}}$ consistently suppresses variance, underscoring that deeper leader–follower negotiation stabilises outcomes.  Third, the gradient of this improvement flattens beyond about $80$ iterations, echoing the scatter plot’s plateau.  These relationships delineate a practical “safe operating zone”: for moderate break rates ($\lambda_{\text{break}}\le 0.003\,\text{s}^{-1}$), roughly $60$ LD-CP iterations keep fairness fluctuations below $0.04$, whereas more severe disturbance regimes require either accelerated solvers to sustain $N_{\text{iter}}\!\ge\!80$ in real time or auxiliary fallback policies.  Taken together, the three visualisations demonstrate that the Stackelberg–MPEC framework not only elevates average fairness but also offers tunable levers—iteration budget and disturbance tolerance—to tailor equity, stability, and computational effort to specific grid-operation requirements.

\section{Conclusion}
This paper presents a unified framework for modeling, optimizing, and validating QKD-enhanced SCADA systems in power grid applications. By integrating quantum key dynamics with multi-timescale control processes, we address the critical challenge of securing power system operations under constrained and uncertain cryptographic resources. Through a Stackelberg game formulation and a novel LD-CP algorithm, we enable fair and adaptive key allocation between transmission and distribution operators. The proposed co-simulation testbed bridges quantum optics, network protocols, and power system control, offering an end-to-end platform for performance evaluation. Experimental results demonstrate significant improvements in control task reliability, frequency stability, and key utilization under dynamic link conditions. Future work will explore the integration of trusted relay networks, cross-layer intrusion detection, and hardware-in-the-loop validation on real-time digital simulators.

\newpage
\vfill

\end{document}